\documentclass{article}[10pt]

\usepackage{jucs2e} 
\usepackage{url}

\usepackage{pifont} 
\usepackage{array} 
\usepackage[percent]{overpic} 
\usepackage{subfig}

\usepackage{csquotes}

\begin{document}

\title{Collaboration Spheres: a Visual Metaphor to Share and Reuse Research Objects}

\author{{\bfseries Mariano Rico\\
   (OEG, Department of Artificial Intelligence, UPM, Spain\\
   mariano.rico@upm.es)
   \and
   {\bfseries Jos\'{e} Manuel G\'{o}mez-P\'{e}rez}\\
   (Expert System Iberia\\
   jmgomez@expertsystem.com)\\
    \and
   {\bfseries Rafael Gonzalez}\\
   (OEG, Department of Artificial Intelligence, UPM, Spain\\
   rgonza@fi.upm.es)\\
   \and
   {\bfseries Aleix Garrido}\\
   (Expert System Iberia\\
   agarrido@expertsystem.com)\\
   \and
   {\bfseries Oscar Corcho}\\
   (OEG, Department of Artificial Intelligence, UPM, Spain\\
   ocorcho@fi.upm.es)\\
}}
\maketitle

\begin{abstract}
Research Objects (ROs) are semantically enhanced aggregations of resources associated to scientific experiments, such as data, provenance of these data, the scientific workflow used to run the experiment, intermediate results, logs and the interpretation of the results. 
As the number of ROs increases, it is becoming difficult to find ROs to be used, reused or re-purposed. New search and retrieval techniques are required to find the most appropriate ROs for a given researcher, paying attention to provide an intuitive user interface.
In this paper we show CollabSpheres, a user interface that provides a new visual metaphor to find ROs by means of a recommendation system that takes advantage of the social aspects of ROs. The experimental evaluation of this tool shows that users perceive high values of usability, user satisfaction, usefulness and ease of use. From the analysis of these results we argue that users perceive the simplicity, intuitiveness and cleanness of this tool, as well as this tool increases collaboration and reuse of research objects.

\end{abstract}


\begin{category}

\begin{enumerate}
\item I.2.1 - Applications and Expert Systems
\item H.3.3 - Information Search and Retrieval
\item H.5.2 - User Interfaces
\end{enumerate}
\end{category}

\section{Introduction} 
The research life cycle in Data-Intensive Science is increasingly oriented towards reusing  experimental results. Now more than ever Science is an incremental task where scientists almost literally stand on shoulders of giants and their previous breakthroughs. From hypothesis and background knowledge to experimental results in the form of scientific datasets and eventually publication as scholarly communications, scientific developments are based on the reuse of previous knowledge and experimentation. As a consequence, scientific publications serve a twofold purpose: i) to communicate a scientific development to the rest of the community and ii) to convince that same community that the claimed results are reliable.

Scientific workflows are well-known means to encode scientific knowledge and experimental know-how. By providing explicit and actionable representations of scientific methods, workflows capture such knowledge and support scientific development in a number of ways, including the validation of experimental results and the development of new experiments based on the reuse and repurpose of existing workflows. Therefore, scientific workflows are valuable scholarly objects themselves, which play an important role for sharing, exchanging, and reusing scientific methods. In fact workflows are usually treated as first-class artifacts for exchanging and transferring actual findings and experimental results, either as part of scholarly articles or as stand-alone objects. This need has been captured by popular public workflow repositories like myExperiment~\cite{Roure2008,Roure2009} and CrowdLabs~\cite{Mates2011}.

Research objects~\cite{Belhajjame2012} (ROs) are means to encapsulate in a single information unit all the information related to a particular experiment. They bundle other digital objects that are relevant for the experiment (papers, datasets, etc.), possibly a workflow, the provenance of the results obtained by its enactment, and annotations that semantically describe all these objects. Therefore, research objects provide a comprehensive view of the experiment, support the publication 
 of experimental results, enable inspection, and contain the information required for the evaluation of the health of a workflow~\cite{Gomez2013b}. 

As today's experiments are not carried out by an isolated researcher but by research groups, normally from different institutions, research objects support communities of users to gain better and more intuitive access to the experimental knowledge contained within. 

Their rich metadata enables the development of sophisticated recommendation methods that exploit such information to make a personalized delivery of relevant research objects to specific scientists.  In combination with such recommendation technology it is possible to build highly expressive user interfaces that simplify the interaction between scientists and research objects. 

We focus on a visual metaphor that combines exploratory search user interfaces and recommendation technologies, aimed at providing a powerful and simplified access to repositories of research objects in scientific communities. This metaphor provides a mechanism to explore, share and reuse research objects and user expertise based on the exploitation of semantic descriptions, relations and similarities between research objects and users in order to provide advanced search functionalities. This type of exploratory search is especially appropriate in domains where social aspects like collaboration and the notion of a community play a relevant role in order to incrementally expand the knowledge assets of such community, like myExperiment.

In addition to this novel visual metaphor, we present an implementation named CollabSpheres, a Web tool based on Linked Data technologies and collaborative filtering-based recommenders. As shown in the evaluation of our approach in a real-life scenario with myExperiment data, the combined use of these technologies increases collaboration and stimulates reuse in scientific communities while reducing the overhead of identifying and retrieving relevant scientific knowledge for a particular purpose.

This paper is structured as follows: section 2 shows related work, section 3 describes the visual metaphor, its architecture, the recommender system, and the types of recommenders used by this tool. Section 4 shows the evaluation of the system, carried out by usability experts over an initial version of the tool, as well as the evaluation carried out by scientists with tool in its final version. Section 5 summarizes and shows conclusions and future work.


\section{Related Work}
The rationale behind the Collaboration Spheres is closely related with circles of collaborative search introduced by Russell-Rose~\cite{Russell-Rose2013}. The definition of the circles of collaborative search provides a holistic view of collaboration in search by defining a three-circle model composed by the inner and intermediate social circles; and the outer circle.  The inner and intermediate circles represent \textbf{explicit} and intentional collaboration between individuals who share some degree of social connectedness, and form the nucleus of collaboration. The outer circle comprises \textbf{implicit}, unintentional collaboration. This is the case in our Collaboration Spheres for the statistical data extracted from the preferences of the whole research community in collaborative filtering. However, our approach goes further, because we show an additional circle of recommendations based on the context (explicit preferences) of the user.

There is also similar work regarding research discovery tools that involve navigating through co-authoring or citation networks represented as graph-like structures. VIVO~\cite{Borner2012} is a semantic Web application built around  a community of universities that aims at enabling collaboration and discovery among scientists across all disciplines. VIVO, among other useful graphical discovery tools such as the Temporal Graph and the Map of Science, provides the Collaboration Network, a graphical representation of a researcher's co-author and co-investigators network extracted from research papers and grants. The researcher is placed in the middle of a circular layout where all the related co-authors and co-investigators nodes are placed around alphabetically (or according to algorithmically-calculated communities), and a colored edge is drawn from the researcher to represent their respective number of collaborations. Compared to our approach, once again, we provide an additional step by allowing users an active participation in the process. 

Microsoft Visual Explorer~\footnote{See \url{http://academic.research.microsoft.com/VisualExplorer}} provides three different graphical representations related to our work: coauthors of a given author, 
degrees of separation between two given authors, 
and citations for a given researcher.
These tools are exclusively focused on the publication network of researchers, providing graph-based visualizations, but no recommendations. We include them here because are popular tools. 

A more recent and less known application is Semantic Scholar~\footnote{See \url{https://www.semanticscholar.org}}, which combines semantic technologies and social aspects in citation networks to enhance access to publications. Although semantics is limited to topic extraction, this tool is similar to the applications derived from CollabSheres, described later on.

ResXplorer~\cite{Vocht2013} is an interactive interface for dealing with research datasets, mostly publications, using Linked Data technologies. Comprises data from 
DBLP(L3S),    
COLINDA,      
DBpedia       
and GeoNames. 
ResXplorer provides a single graph that consolidates all the information in the shape of a radial layout where each node differs in their shape and color, used to distinguish between the different node types (currently agents, conferences, proceedings and documents); and their size, determined by its novelty and relation to the context, attracting thus the user attention to the most relevant. We consider that ResXplorer is relevant to our work because uses linked data technologies, although our tool deals with research objects and uses different recommenders.  

Finally, this work is also related to SmallWorlds~\cite{Gretarsson2010}, a  visual interactive tool which generates item recommendations based on a combination of Facebook user profile data and user interactions, providing a graph-based interface. This graph is composed by user nodes linked through their shared items (such as books, movies, music, etc.) The graph is arranged as a multi-layered radial layout with 5 layers: layer 1 represents the active user's node, layer 2 contains the items that belong to the active user, layer 3 shows similar friends who have items in common with the active user, layer 4 shows the candidate recommendation set (composed by items that are not in the active user's profile but are liked by friends in layer 3), and layer 5 contains friends who have no items in common with the active user abut they have with friends in layer 3. The positions inside the layers and the size of each node are calculated algorithmically using similarity functions combined with a recommendation algorithm that uses the tastes of the active user and her similar friends. Our approach provides a layered view conceptually similar to the one provided by SmallWorlds, but simplified (instead of 5 layers we provide 4). Also we share with this application the social and collaborative aim. However, our recommender system takes advantage of linked data technologies, as ResXplorer does (but oriented to exploit research objects instead of publications), and we have different recommenders. 

\section{Collaboration Spheres as a Visual Metaphor}
There are different ways in which users can express a query in order to retrieve content from a repository. Well-known approaches include, among others, faceted search[1], where the user selects the most relevant features from a predefined set in order to constrain the search space, and free text query interfaces, which rely on natural language processing technologies to parse and match the query against the overall content. However, it is usually the case that users lack the precise knowledge about the exact features of the information to be retrieved or the skills required to express them in specific query formalisms.

The difficulty is considerably lower when, instead of formulating a query, users are enabled to search by example, exploiting the potential similarities between such examples and the results. We follow the same principle behind the use of examples e.g. in education, in order to facilitate the assimilation of complex concepts by students. By selecting a number of representative exemplars users can express the properties that must be observed in the expected results without explicitly or formally describing such properties. This query-by-example method relieves users from the task of formulating potentially complex queries, delegating such complexity to the underlying system.

Additionally, this approach allows users to explore the search space through the context of interest described by the aggregated properties of the selected exemplars. For example, by putting together different ROs dealing with cardiovascular diseases and diabetes and analyzing the relatedness of this context of interest against the ROs contained in the repository it is possible to establish a connection between these disorders and the metabolic syndrome. This kind of exploratory search is specially aimed towards gaining new insights on existing information and discovering relations between them that were not previously explicit.

The collaboration spheres intent to provide a mechanism to improve, \textbf{share} and \textbf{reuse} ROs and \textbf{user experience} based on the exploitation of semantic descriptions, relations and similarities between ROs and users in order to support advanced search mechanisms. The search activity achieved by using the visualization of those similarities has a very strong social analysis aspect and the implemented work is based on collaborative filtering and content-base recommendations provided by the Recommender Service described in the next sections. 

The visual metaphor is based on spheres centered around a central point that represents the user. This metaphor intends to be simple and user oriented providing a link between the Recommender Service, the ROs meta-data, and the user interaction.

The CollabSpheres web application is described in the next section. The implementation code is available at GitHub~\footnote{See \url{http://github.com/wf4ever/Collaboration-spheres}}.

\subsection{Description of the user interface}
In this section we describe the user interface of CollabSpheres, the web application that mediates between the end user (a given scientist) and the recommendation system. We describe the interface by means of the application available at \url{http://sandbox.wf4ever-project.org/collab2/index.html?id=http://www.myexperiment.org/users/18}. Notice that the specific scientist is the \texttt{id} parameter in the URL, in this example the \texttt{id} correspond to a specific myExperiment user named "Marco Ross". You can change the scientist by changing its myExperiment \texttt{id} in the URL.

\begin{figure*}
    \centering
    \subfloat[Initial state of the CollabSpheres Web application. For the person in the central circle (labeled \ding{182}) the user gets an initial recommendation in the white circle (labeled \ding{183}). Clicking on any element in the lists of friends and ROs (labeled \ding{184}) shows a detailed description of the element in the lower right box (labeled \ding{185}).]{
      \label{fig:GUI_initial}
      \begin{overpic}[width=5.4in]{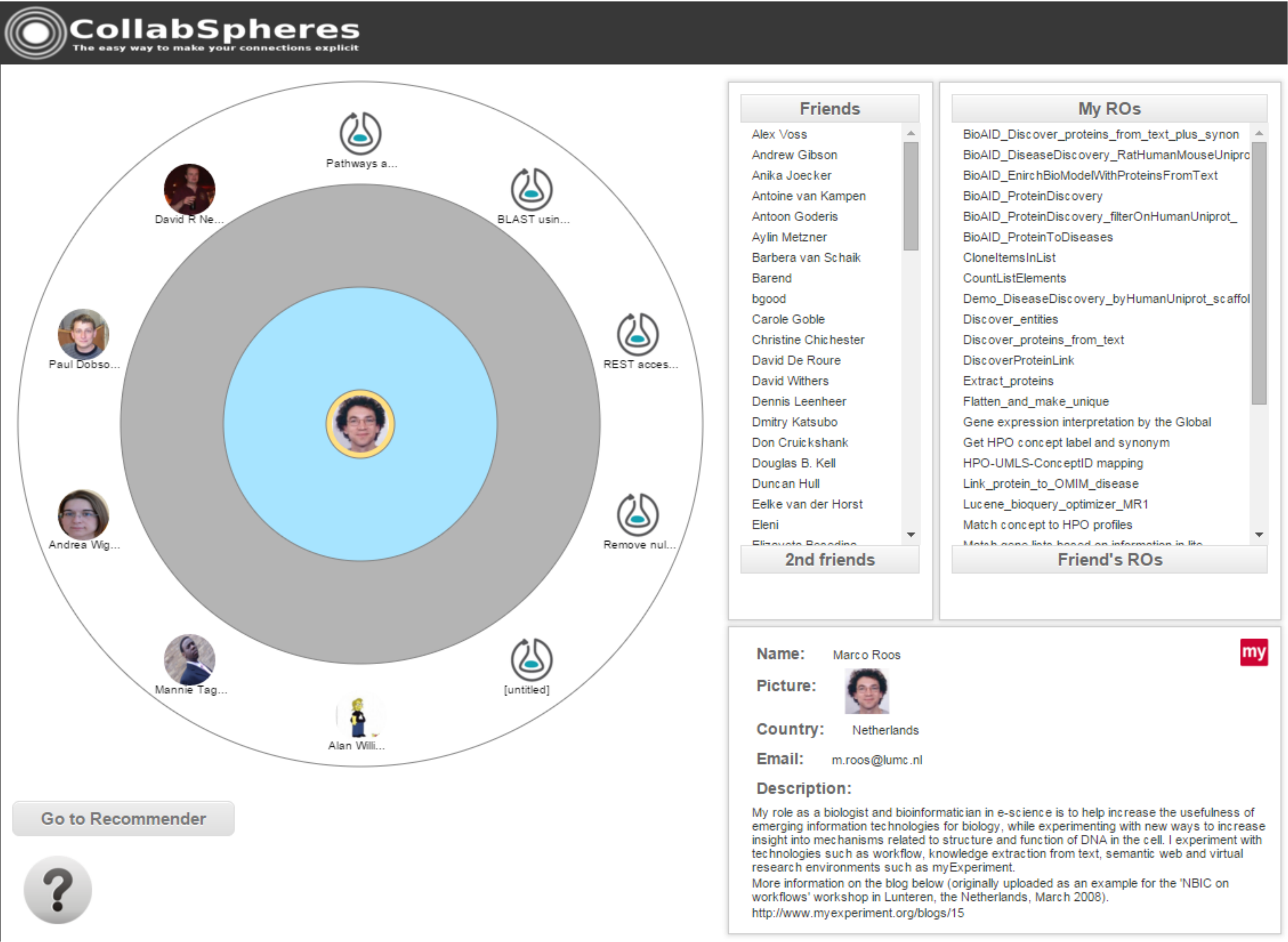}
        \put (29,38) {\huge \ding{182}} 
        \put (31,15) {\huge \ding{183}} 
        \put (71,50) {\huge \ding{184}} 
        \put (80,15) {\huge \ding{185}} 
      \end{overpic}}
    \hspace{0.2cm}
    \subfloat[A friend's RO has been dropped in the blue circle (labeled \ding{186}). System recommendations are computed and located in the gray circle (labeled \ding{187}). The details of the recommendation can be displayed pressing the "Go to Recommender" button (labeled \ding{188}). The recommendations for this case are shown in figure~\ref{fig:GUI_recommender}.]{
      \label{fig:GUI_final} 
      \begin{overpic}[width=5.4in]{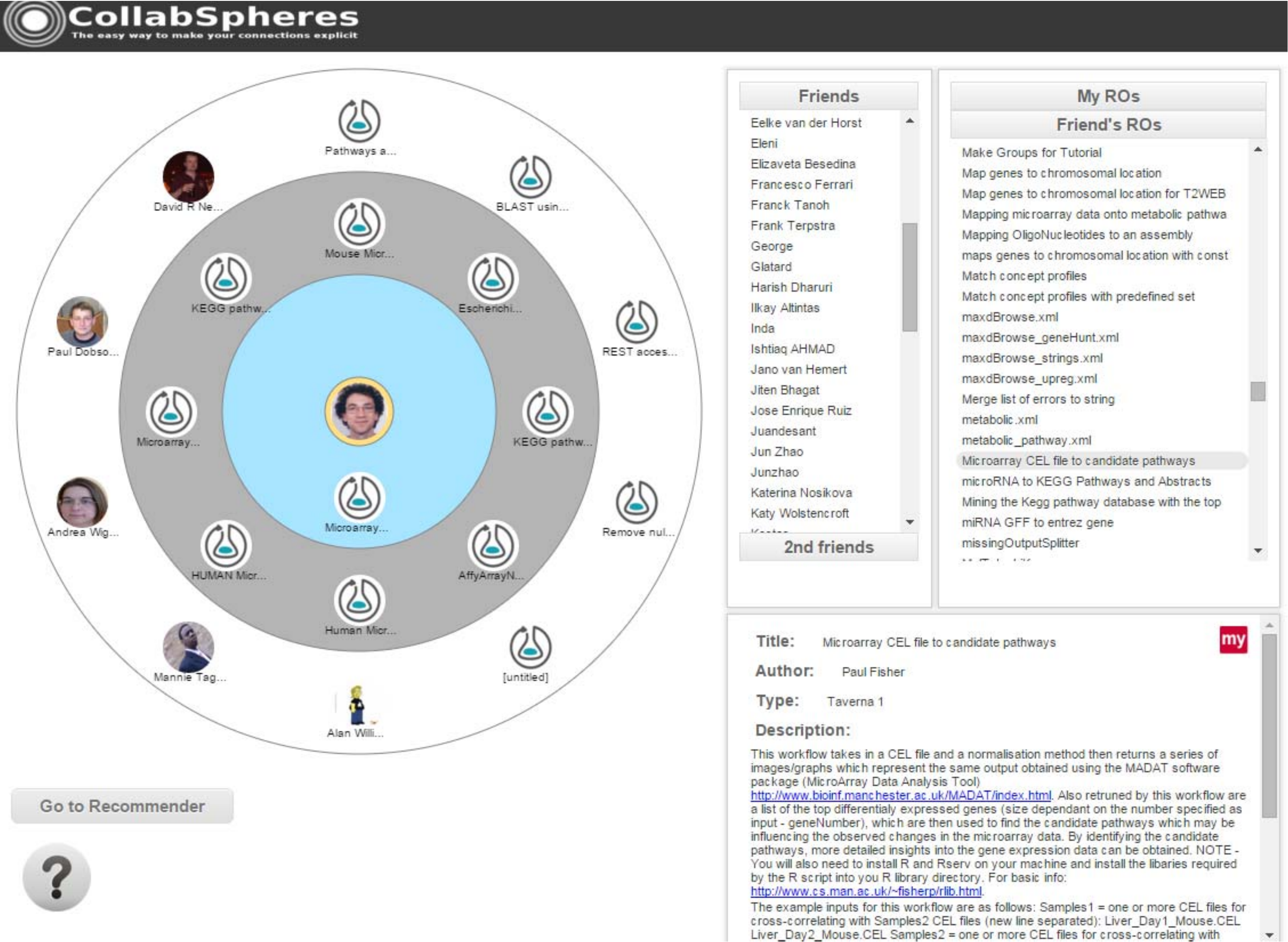}
        \put (32,39) {\huge \ding{186}} 
        \put (31,26) {\huge \ding{187}} 
        \put (18,9)  {\huge \ding{188}} 
        \put (75,37) {\vector(-1,-0){45}\thicklines}  
    \end{overpic}}
    \caption{The User Interface of the CollabSpheres web application.}
\end{figure*}

Figure~\ref{fig:GUI_initial} shows the initial state of the Collaboration Spheres web application provided by the aforementioned URL. There are three concentric circles around Marco Ross (labeled \ding{182} in the figure). The most external circle (white circle labeled \ding{183}) contains the basic recommendations, based on the ROs of Marco and his friends. The lists in the upper right side of the interface (labeled \ding{184}) show a) friends of Marco, b) friends of Marco's friends, denoted by '2nd friends', c) Marco's ROs, denoted by "My ROs", and d) ROs of Marcos's friends, denoted by "Friend's ROs". When the users clicks in the central circle (on the picture of Marco), or on any person or RO in the lists, a detailed description (provided by myExperiment) is shown in the lower right box of the user interface (labeled by \ding{185} in the figure).

Users can drag\&drop people or ROs from the lists to the proximity of the person in the central circle in order to enhance the basic recommendation. Figure~\ref{fig:GUI_final} shows the effect of dragging a specific RO into the proximity of the central circle (see \ding{186} in the figure). In this example we have dragged one element in the "friend's ROs" list, specifically "Microarray CEL file to candidate pathways". The system computes the recommendations and place the best ranked ROs in the gray circle (see label \ding{187} in figure~\ref{fig:GUI_final}). 

As the blue circle (\ding{186}) is closer to the central person than the grey one (\ding{187}) or the white one (\ding{183}), the user perceive (correctly) that recommendations in the grey circle (\ding{187}) are better than recommendations in the white circle.

Users can drag\&drop more ROs and people to get a better recommendation. Users also can remove ROs (or people) previously dropped by clicking on the RO (or person) with the right button. Recommendations derived from this RO (or person) will be removed as well, and recommendations are recomputed. Users can read a report on the last recommendation clicking the 'Go to Recommender' button (labeled \ding{188}).


\begin{figure*}
    \centering
    \includegraphics[width=4.0in]{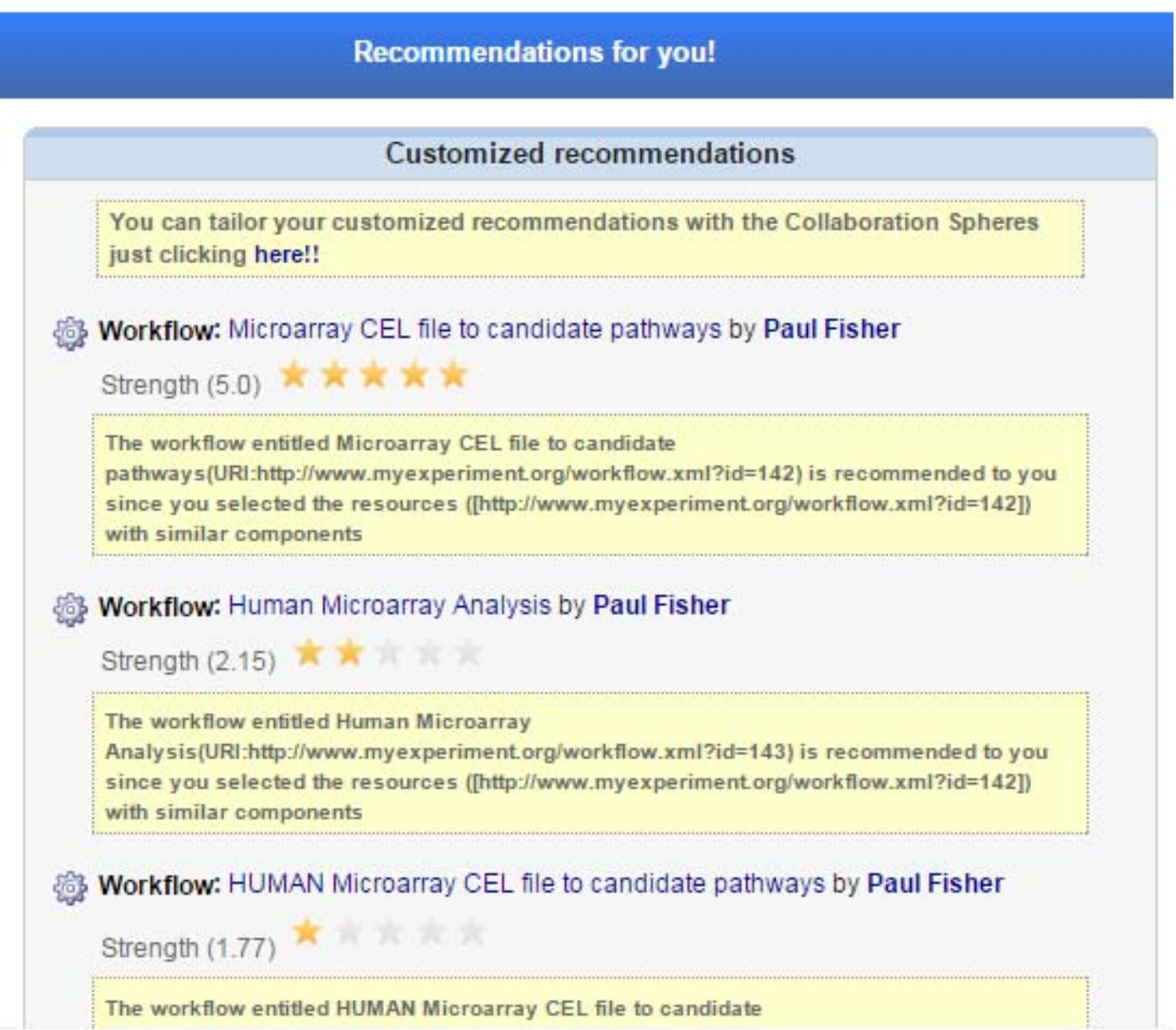}
    \caption{Recommendation explanation for the example in figure~\ref{fig:GUI_final}.}
    \label{fig:GUI_recommender}    
\end{figure*}

Figure~\ref{fig:GUI_recommender} shows the recommendation report obtained after the actions described in figure~\ref{fig:GUI_final}. The user can read here the explanation of the reasons, and the ranking (strength, in both numerical and starred ways), for all the recommended items. Notice that the gray circle only shows the top most relevant recommendations due to space restrictions, but the report shows all the recommendations computed by the system.

\subsection{Architecture}\label{sec:achitecture}
The CollabSpheres web application relays on the Linked Data~\cite{Bizer2009} and a solid server architecture for digital preservation
compliant with standards like the Open Archival Information System (OAIS) reference model~\cite{OASIS2009}(ISO 14721-2002 standard). With this architecture we combine RO lifecycle management, social networks and digital libraries. 

\begin{figure*}
    \centering
    \includegraphics[width=5.0in]{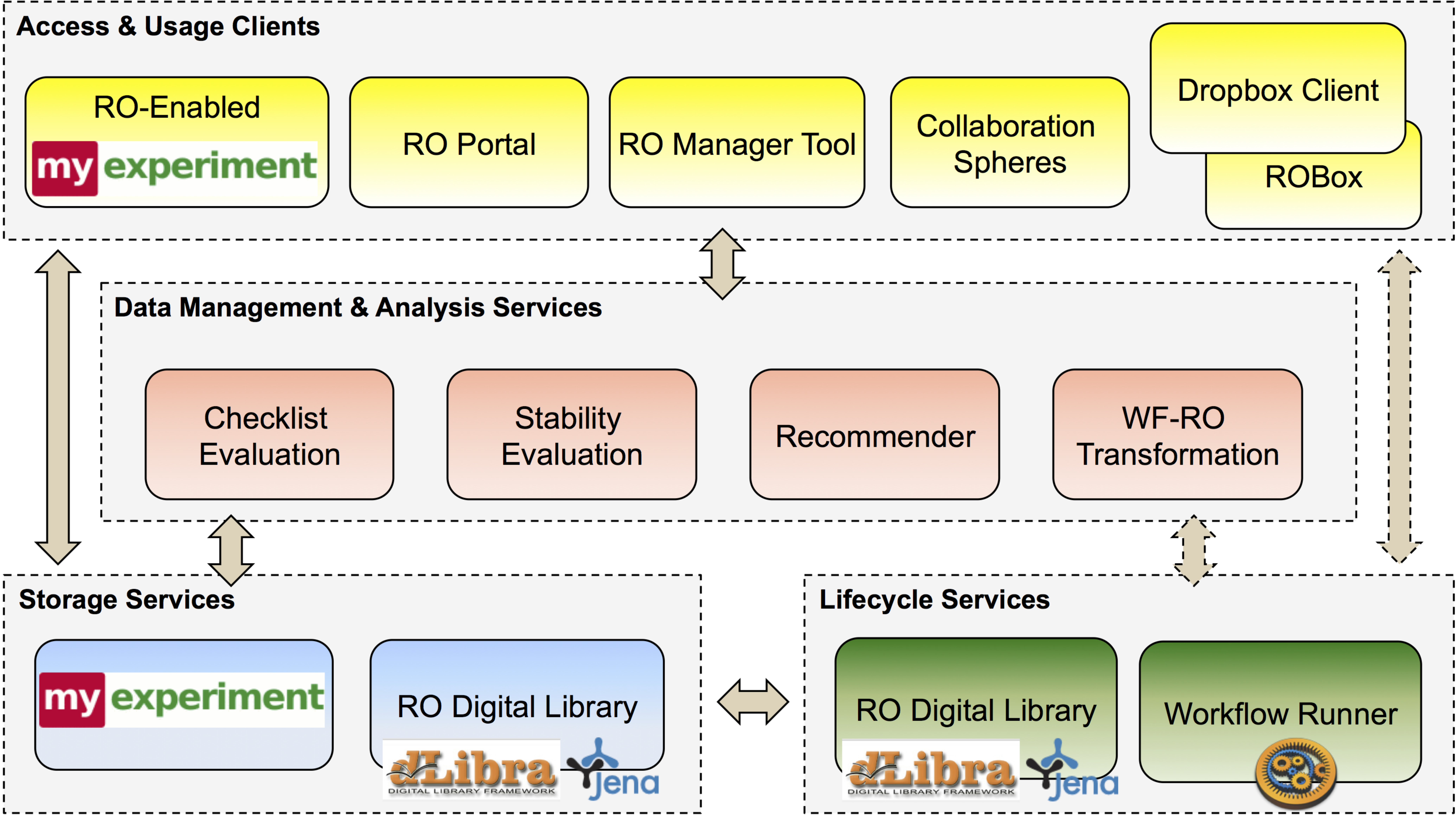}
    \caption{Wf4Ever architecture. The CollabSpheres web application is located in the top layer. Rounded rectangles denote services and bold texts denote APIs.}
    \label{fig:Wf4EverArchitecture}    
\end{figure*}

This application is on top of the Wf4Ever ecosystem~\cite{Page2012}. Figure~\ref{fig:Wf4EverArchitecture} shows the main components of this infrastructure: three layers of functionality in which rounded rectangles denote services and bold texts denote  APIs. Into the top layer, labeled "Access \& Usage Clients", you can see the rounded rectangle entitled "Collaboration Spheres". This layer has access to the API "Data Management \& Analysis Services", in which the recommender is located, as well as the access to the lower layers: "Storage Services" (where myExperiment data are located) and "Lifecycle Services" (where the ROs interact with workflows).

\begin{figure*}
    \centering
    \includegraphics[width=5.0in]{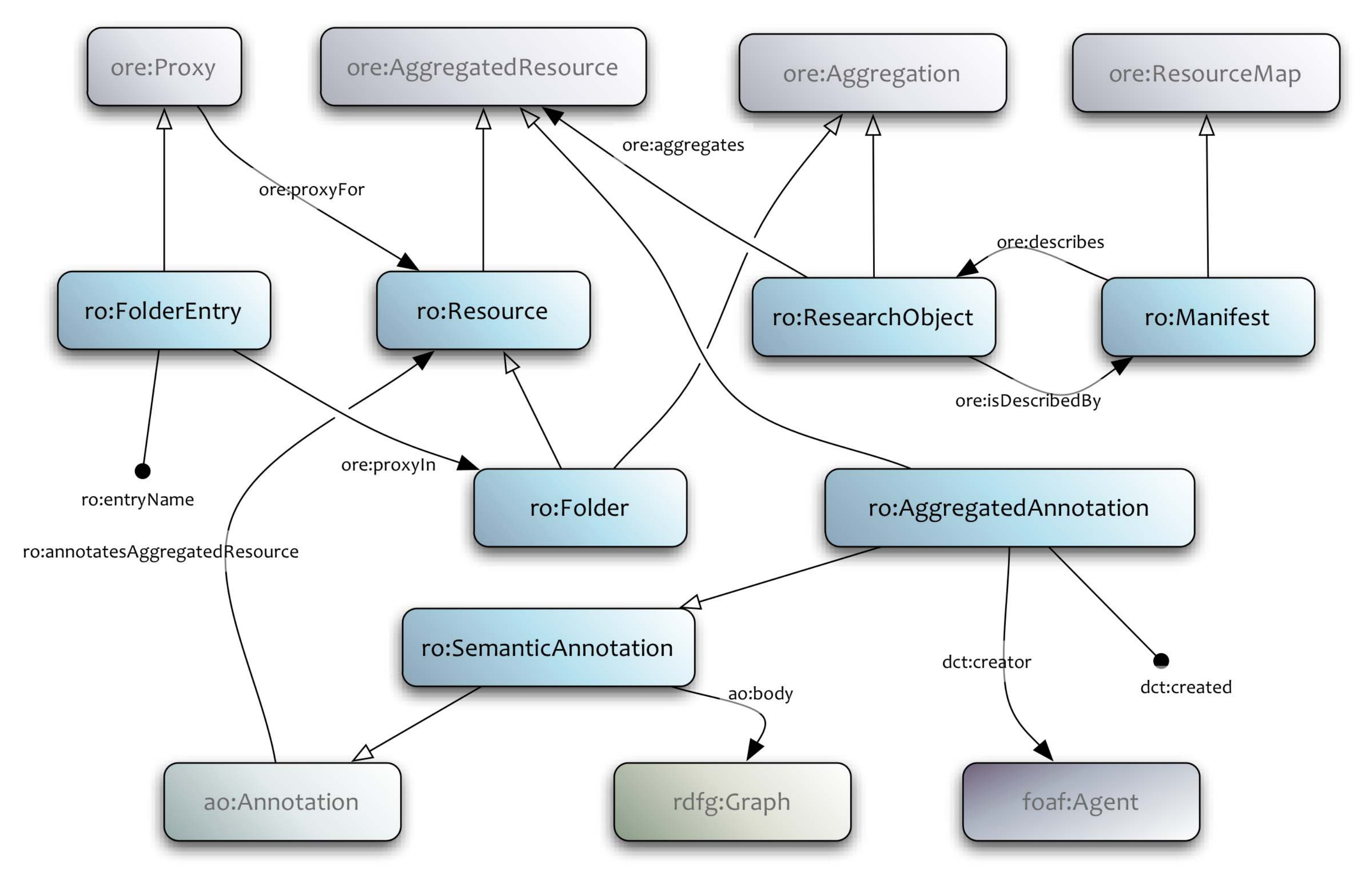}
    \caption{The Research Object ontology model, showing the relation between the \texttt{ro:ResearchObject}, other ontology's elements and elements from other ontologies.}
    \label{fig:research_object_model}    
\end{figure*}

In this architecture, Research Objects are modeled using the Research Objects Ontology (denoted by the \texttt{ro} prefix)~\footnote{See \url{http://www.wf4ever-project.org/vocab/ro\#}}. Figure~\ref{fig:research_object_model} shows the main concepts related to the \texttt{ro:ResearchObject}. This OWL ontology extends the OAI-ORE ontology~\footnote{See \url{http://www.openarchives.org/ore/}} (prefix \texttt{ore} in the figure) and includes other common ontologies like the Annotation Ontology (prefix \texttt{ao} in the figure) or the FOAF ontology.

This architecture follows a RESTful approach for the design of its service APIs,
enabling the development of relatively lightweight client software. All services are exposed as a set of inter-linked resources over HTTP, and a client ``browses'' through the links to find other resources. The contract between client and server is the set of media and link types that the server returns as available, and which the client must ``understand''  to be able to proceed and retrieve resources from the server and navigate to further linked resources. To some degree this allows a client to work with only a partial implementation of the full set of ``calls'' offered by the services. That is, the client can ignore some link and media types if the software designer believes it is appropriate; there is no need to fully match all the operations offered by the service. 

This also means that possible state transitions are controlled by the server, though the actual movement from one state to another is taken by the client. Since resources and the links between resources can be manipulated at any time on the server, the client can only rely on what it has retrieved from the current resource and the links therein. That is, there must be a working assumption that every time a resource is retrieved it may be different.

While this adds considerable complexity to the design of the API, it removes much complexity from client design since it follows a path given by the server. It is somewhat similar to a person using a browser, in which the person follows and interprets web pages and links.

\begin{figure*}
    \centering
    \includegraphics[width=5.0in]{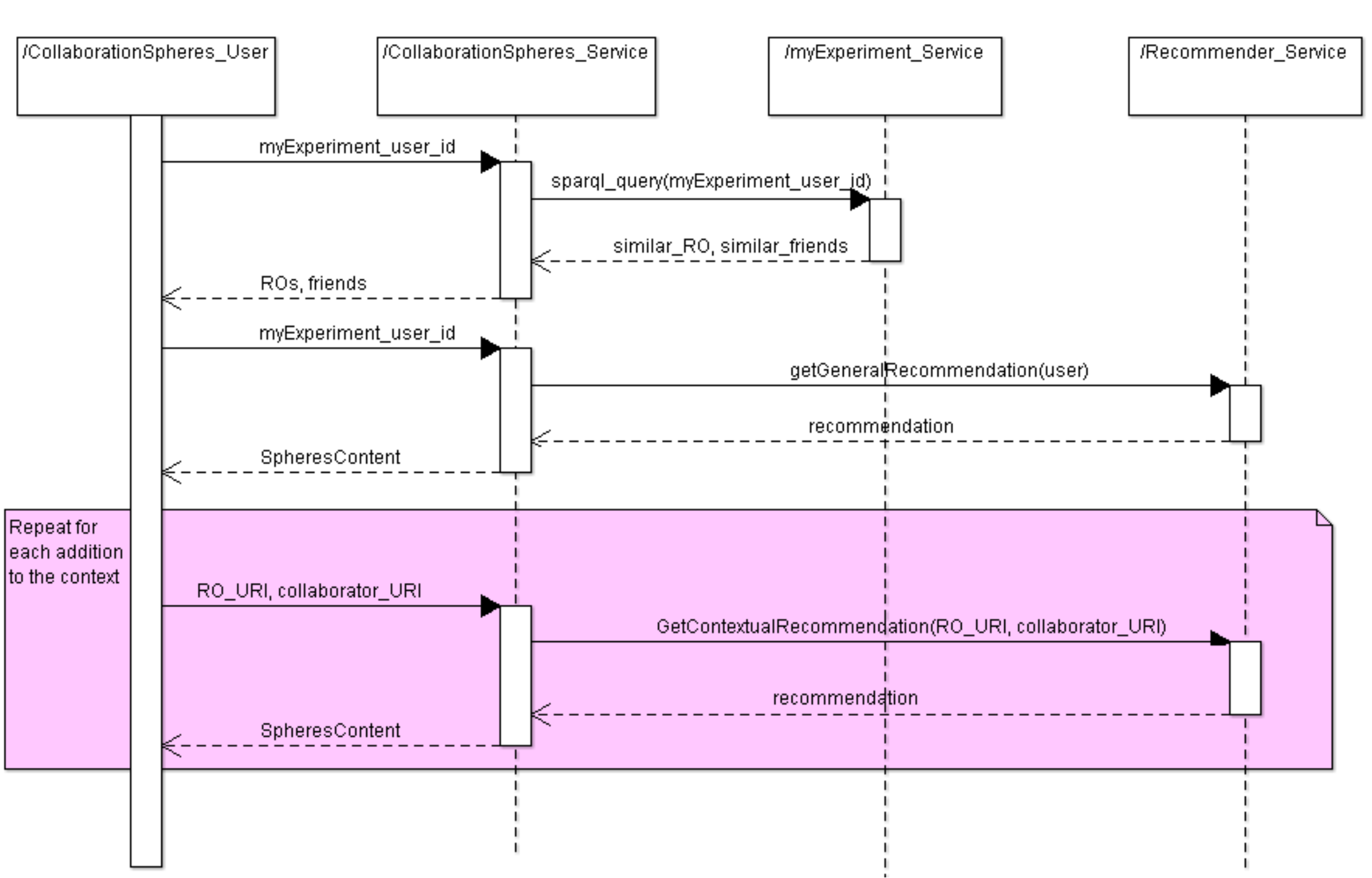}
    \caption{Sequence diagram between the CollabSheres user interface and the recommender system.}
    \label{fig:CollabSpheres_and_Recommeder_sequeceDiag}    
\end{figure*}

Figure~\ref{fig:CollabSpheres_and_Recommeder_sequeceDiag} shows an example of the sequence of messages between the CollabSpheres user interface and the recommender system. Notice that the communication is between services, due to the aforementioned architecture of the system. The linked data approach can be seen in the SPARQL query made to the myExperiment repository (function \texttt{sparql\_query}), as well as collaborators are identified by IRIs.

\subsection{The Research Object Recommender Service}
The Recommender Service brings useful hints to the researcher in a proactive manner providing, without prior request by the user, practical suggestions of scientific data and results in the shape of: 
\begin{itemize}
\item ROs as a whole.
\item  Resources that compose ROs. The system may suggest resources that might be useful additions/alternatives to the ones already aggregated by the ROs that the user is currently using or creating.
\end{itemize}

The advantage of the recommendation activity over search mechanisms in the scenario of scientific content discovery is that recommender systems are not passive and that they do not presume that the user has the notion of the existence of possibly relevant scientific content. The Recommender Service performs such recommendation activity using novel multi-faceted recommendation techniques mixing community-based and advanced demography/content-based recommendations along with collaborative filtering techniques.

Besides these recommendation activities, we also considered social aspects of
research that were not considered in the initial catalog of requirements but have turned out to be of great importance. As a consequence, to the set of items that the Recommender Service considers in its recommendations we also include the
recommendations of other researchers on the basis of the user's social network.

This service handles the information of 6914 myExperiment users, 1759 workflows, 865 scientific content related files, 370 packs (myExperiment equivalent of ROs). With such information it is able to provide 6345 recommendations (177 using the Collaborative Filtering recommender, 4639 using the Keyword Content-Based recommender and 1799 using the Social Network recommender); and the Inference Engine using such recommendations has inferred 225 pack recommendations. Finally, from the 6914 myExperiment users, 608 have received at least one recommendation.

The source code is publicly available at GitHub~\footnote{See \url{http://github.com/wf4ever/epnoi} for the core functionality and \url{http://github.com/wf4ever/epnoiServer} for the server based in the Grizzly application container}. The details of its interface, data format, configuration, deployment, etc. are available at the Wf4Ever project public wiki pages~\footnote{See \url{http://www.wf4ever-project.org/wiki/display/docs/Recommender+Service}}.

\subsection{Recommenders}
\begin{figure*}
    \centering
    \includegraphics[width=3.5in]{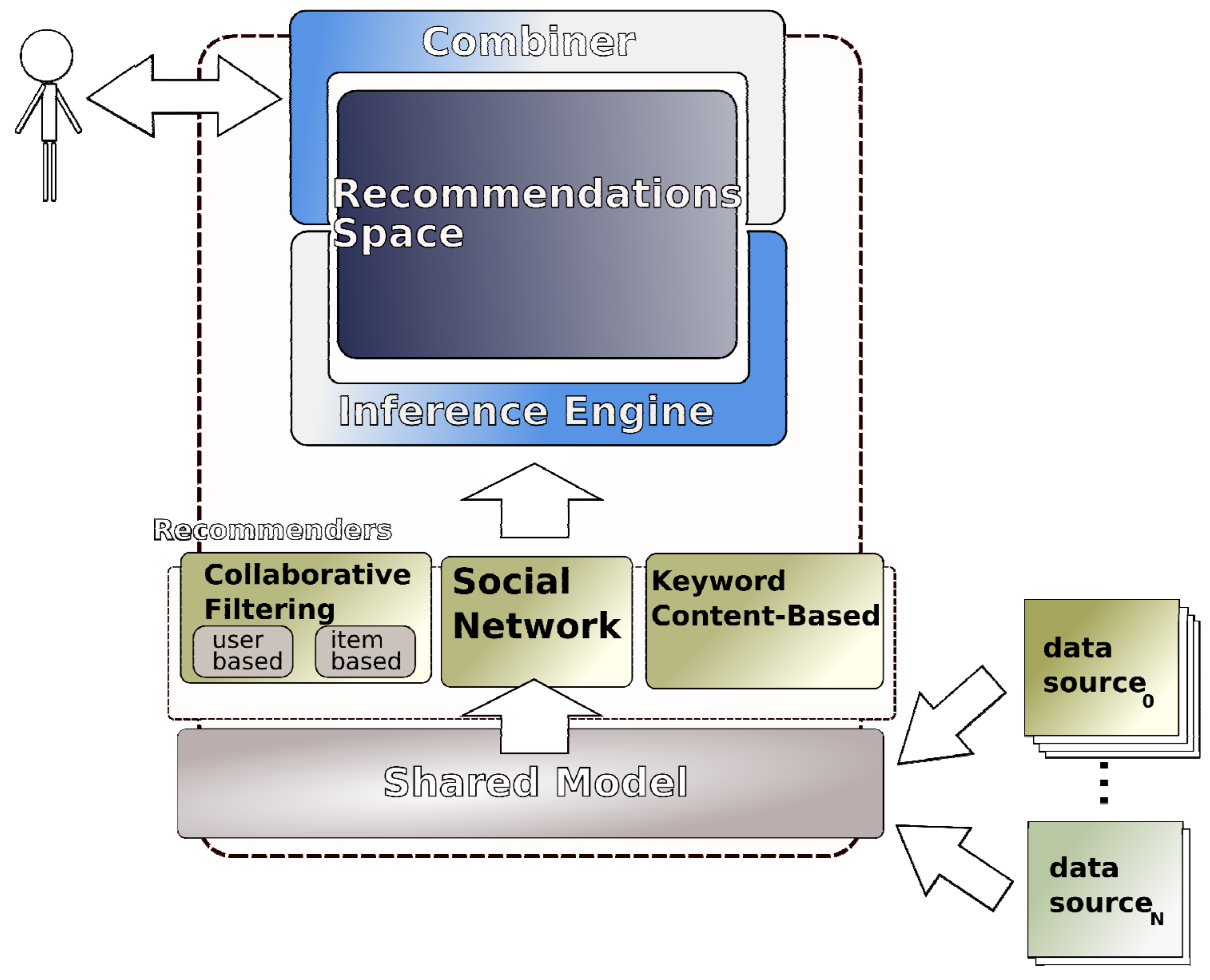}
    \caption{Recommender Service architecture.}
    \label{fig:Recommender_architecture}    
\end{figure*}

As shown in figure~\ref{fig:Recommender_architecture}, the Recommender Service uses a variety of recommenders, each of them with its own recommendation algorithms and parameterization. That is, we consider the case of more than one recommender using the same recommendation algorithms with different parameters. The recommenders types currently deployed are 

\begin{enumerate}
\item \textbf{Collaborative Filtering recommender}. Predicts user's affinity for items on the basis of the ratings that other users have made to these items in the past~\cite{Hill1995, Shardanand1995}. Therefore, the recommendation is based on finding people with similar tastes to the user (or items with similar rating patterns as the one that the user has rated) by means of its past ratings; and by means of their ratings extrapolate the user future ratings. User information in a collaborative system consists of a vector of items and their associated ratings; finding similar users translates into finding similar vectors. 

The algorithm computed for each user has these two steps: (1) User neighborhood determination, and (2) Inference of new ratings values and creation of recommendations.

With the inclusion of the Collaborative Filtering recommender we address the discoverable RO requirement and, more importantly, the reputation requirement from the catalog of requirements that were identified for the Recommender Service.

The main disadvantages of collaborative filtering are due its dependence on large historical data set for getting good quality results.

\item \textbf{Social Network recommender}. Provides recommendations of users on the basis of their interactions with other users; both from a social perspective (i.e. interactions with other users that have been previously labeled as friends by the user) and authorship network perspective (i.e. other user's that have co-authored scientific content in the past). This recommender uses a three steps algorithm: (1) Creation of the social network graph, (2) Computation of similarity measurement (Interaction Similarity algorithm), and (3) Creation of users recommendations.

With the inclusion of the Social Network recommender we address the social aspects of users that where not identified as a requirement for the Recommender Service, but identified as an important factor.

\item \textbf{Keyword Content-based recommender}. Content-based recommender systems~\cite{Bechhofer2010, Lang1995} make use of information retrieval and filtering techniques. A recommender of this type tries to infer users future items of interest on the basis of the features of the objects that the users rated in the past. These object features are items of interest, e.g. keywords that define the object or a summary of its content.

Content-based techniques have similar advantages to collaborative filtering approaches (without the ability of detecting cross-genre niches), and they do not exhibit the new item problem. Nonetheless, they still rely in a large historical data set.

The main disadvantage of content-based recommendation algorithms is the new user-handling problem. When an user approaches to the system for the first time, the system still do not have a well-formed user's profile. Nevertheless, this problem is less acute that in the case of the collaborative filtering techniques, as this technique does not rely on long-time statistical information; it just needs that the user provides a small set of keywords that represent its interests.

The algorithm that the Keyword Content-based recommender can be described in these three steps: (1) Creation of the description of the content of items, (2) Creation of the user profile of interest, and (3) Matching process. 
\end{enumerate}

\subsection{Recommendations Combiner}
Recommender systems are inherently vertical and are configured to provide recommendations in a single and specific domain. We need ways for tailoring specific recommendations in terms of each research community that in the future wishes to make use of the recommender system. We address this combining the recommendations obtained by the recommendation algorithms described previously. We implemented the state-of-the-art in hybrid recommendation systems~\cite{Burke2002}, although the current online version only provides a straight-weighted combination.

\section{Evaluation of the system}
In order to evaluate the Collaboration Spheres user interface, we carried out two experiments. The first one was oriented to usability experts, aimed at detecting usability problems in the early stages of the prototype. We implemented most of their comments to get an usability-enhanced tool. With this new tool we made a second experiment oriented to final users (scientists), aimed at obtaining several quality related measurements, specifically usability, satisfaction, usefulness and ease of use. Both experiments are described in the next sections.
\subsection{Early detection: The expert's evaluation of usability}
Five usability experts were recruited from our academic institution. They were requested to evaluate independently early-prototype versions of the system web application. During the evaluation session each evaluators used the applications several times, from different starting points, and inspected the interactive elements. They had to answer questions related to usability ``eight golden rules''~\cite{Shneiderman2005}, shown in Table~\ref{table:eightGoldenRules}.

\begin{table*}
    \small  
    \caption{Usability ``Eight golden rules''.}
    \label{table:eightGoldenRules}
    \centering
    \begin{tabular}{|>{\centering}m{0.5cm}|m{2cm}|m{9.4cm}|} 
        \hline
        \textbf{ID} & \textbf{Rule} &  \textbf{Description}
        \tabularnewline
        \hline
        1 & Consistency & There must be consistency in the actions, terminology (messages, menus and help windows) and graphics (colors, layout, and fonts).
        \tabularnewline
        \hline
        2 & Universal usability & Each user has a need; therefore the system must provide some facilities in order to transform contents. Not only impaired
users, but differences between beginners-experts (beginners need explanations, the experts need shortcuts), or age ranges.
        \tabularnewline
        \hline
        3 & Informative feedback & Each action in the system must produce a feedback. For common actions not very important, the answer must be small, but infrequent actions or important must produce a higher response.
        \tabularnewline
        \hline
        4 & Dialogs & Dialogs must be designed to finalize something. Sequences of actions must be organized in groups with a start, middle, and
final. For example, the concept of cart in web applications, with visualization for finished stages and pending stages.
        \tabularnewline
        \hline
        5 & Error prevention & The system must avoid that users make mistakes, but if the error is produced, the system must provide with a solution simple, constructive and specific.
        \tabularnewline
        \hline
        6 & Undo & Allow users to undo actions in an easy way. Everything should be undo-able.
        \tabularnewline
        \hline
        7 & Locus internal control & Support for the locus internal control. The expert users must have the sensation of controlling the tool. Users must start
actions, not only respond to them.
         \tabularnewline
        \hline
         8 & Memory load & Diminish the memory load in the short-term. Avoid multiple windows, codes, or complex sequences.
        \tabularnewline
        \hline
    \end{tabular}
\end{table*}

The questionnaire (see table~\ref{table:questUsab}) comprised ten questions from Nielsen \cite{NIelsen1994uim}. This is a Likert scale-based questionnaire, i.e. ``one based on forced choice questions, where a statement is made and the respondent then indicates the degree of agreement or disagreement with the statement on a 5 (or 7) point scale''~\cite{Brook1996sus}, with integer possible values for the agreement level concerning the sentences in the range from 1 (hard) to 5 (or 7) (easy). Besides the choice, each question had an optional comments field.  Additionally, each evaluator provided us with a list of recommendations to improve usability. These comments and recommendations were analyzed, and most of them were implemented in the final version of the application.

\begin{table*}[t]
    \small  
    \caption{Questionnaire for usability experts. Experts had to indicate their agreement level with these sentences.}
    \label{table:questUsab}
    \centering
    \begin{tabular}{|>{\centering}m{0.3cm}|
                     >{\raggedright}m{11cm}|}  
        \hline
        \textbf{ID} & \textbf{Question}
        \tabularnewline  \hline
        1 & \textit{Visibility of system status}\\ 
The system keeps users informed about what is going on, through appropriate feedback within reasonable time.   
        \tabularnewline
        \hline
        2 & \textit{Match between system and the real-world}\\
The system speaks the user language, with words, phrases and concepts familiar to the user, rather than system-oriented terms. Follow real world conventions, making information appear in a natural and logical order.
        \tabularnewline
        \hline
        3 & \textit{User control and freedom}\\
The system provides a clearly marked ``emergency exit'' to leave the unwanted state without having to go through an extended dialogue. There is support for undo and redo actions.
        \tabularnewline
        \hline
        4 & \textit{Consistency and standards}\\
Users do not have to wonder whether different words, situations, or actions mean the same thing. Platform conventions are followed.
        \tabularnewline
        \hline
        5 & \textit{Error prevention}\\
The system eliminates error-prone conditions or check for them and present users with a confirmation option before they commit to the action.
        \tabularnewline
        \hline
        6 & \textit{Recognition rather than recall}\\
The user's memory load is minimized by making objects, actions, and options visible. The user does not have to remember information from one part of the dialogue to another. Instructions for use of the system are visible or easily retrievable whenever appropriate.
        \tabularnewline
        \hline
        7 & \textit{Flexibility and efficiency of use}\\
There are accelerators ``unseen by the novice user'' intended to speed up the interaction for the expert user such that the system can cater to both inexperienced and experienced users. Allow users to tailor frequent actions.
        \tabularnewline
        \hline
        8 & \textit{Aesthetic and minimalist design}\\
Dialogues do not contain information which is irrelevant or rarely needed. Every extra unit of information in a dialogue competes with the relevant units of information and diminishes their relative visibility.
        \tabularnewline
        \hline
         9 & \textit{Help users recognize, diagnose, and recover from errors}\\
Error messages are expressed in plain language (no codes), precisely indicate the problem, and constructively suggest a solution.
        \tabularnewline
        \hline
        10 & \textit{Help and documentation}\\
The system provides help and documentation. Any such information is easy to search, focused on the user's task, list concrete steps to be carried out, and is not too large.
        \tabularnewline
        \hline
    \end{tabular}
\end{table*}

The average usability value was 5.6 (in the range [1, 7]), with std. dev. 1.1.
Question Q3 (``User control and freedom'') had low values due to the lack of undo/redo features. 


The qualitative recommendations from the evaluators and their relationship to the ``eight golden rules'' are summarized in table~\ref{table:usabRecomend}. Some recommendations were added to the final version of the web application (e.g. Rec1, Rec2), but other could not be followed because the current version of the  interface API does not support them (e.g. Rec3) or because they require deep modifications in the application components (e.g. Rec4).

\begin{table*}
    \small  
    \caption{Aggregated recommendations provided by usability experts.}
    \label{table:usabRecomend}
    \centering
    \begin{tabular*}{12cm}      {|>{\centering}m{0.7cm}
                                 |>{\raggedright}m{3.8cm}
                                 |>{\centering}m{0.4cm}%
                                 |>{\raggedright}m{6.7cm}|}
        \hline
        {\rotatebox{90}{\textbf{Recommendation ID }}} & \textbf{Recommendation} &  {\rotatebox{90}{\textbf{G-Rules involved }}} & \textbf{Solution implemented in the final version of the application}
        \tabularnewline
        \hline
        Rec1 & Sometimes the system delays and there is no feedback & \#1 \#3 & The application added a dynamic icon with text ``loading'' to provide feedback to the user during delays.
        \tabularnewline
        \hline
        Rec2 & Links to people or ROs to avoid codes or identifiers & \#5 \#6 \#9 & Links were provided in Recommender messages and in the description of people/ROs.
        \tabularnewline
        \hline
        Rec3 & Would be useful to have inter-spheres drag\&drop & \#2 \#6 \#8 & This recommendation requires deep modifications of the graphics libraries used by the application.
        \tabularnewline
        \hline
        Rec4 & I miss a global undo & \#3 \#9 & This recommendation requires deep modifications of the application.
        \tabularnewline
        \hline
    \end{tabular*}
\end{table*}

\subsection{End-user evaluation}

In order to measure the quality of the CollabSpheres web application, a set of 15 participants were selected from our institutions. Participants played the role of a given researcher in the myExperiment platform (specifically, \url{http://www.myexperiment.org/users/18}) and they had to complete a use case in which most of the functionality described in the previous sections was covered. Although there was no maximum time to finish the use case, most users required around 30 minutes. The participants loaded this role in CollabSpheres through this URL: \url{http://sandbox.wf4ever-project.org/collab2/index.html?id=http://www.myexperiment.org/users/18} (notice that the specific myExperiment user is the \texttt{id} parameter in the URL).

Once finished, a detailed questionnaire was presented to each participant. This questionnaire (translated to English), as well as the aforementioned use case, is available in the \textit{Resources} section of the page at \url{http://sites.google.com/site/spheresquestionnaire}. This questionnaire has 50 Likert questions and 4 free text questions, evaluating four different aspects of our tool:
\begin{enumerate}
\item Usability of the web application. Comprises 13 questions in the range 1 to 5, divided in 3 groups: 4 questions related to \textit{learning}, with ID from U-L1 to U-L4; 5 questions related to \textit{adapting to the user}, from U-A1 to U-A5;  and 4 questions related to \textit{feedback and errors}, from U-F1 to U-F4). This questionnaire part finishes with 2 free text questions requesting the most negative and positive aspects of the tool. 
\item User's satisfaction concerning the User Interface of the application. It has 25 questions in the range from 1 to 7, divided in 5 blocks: 6 questions for \textit{Overall reaction to the application} (S-Q1 to S-Q6), 3 for \textit{Screen} (from S-P1 to S-P3), 5 for \textit{Terminology and system information} (from S-I1 to S-I5), 6 for \textit{Learning} (from S-L1 to S-L6) and 5 for \textit{System Capabilities} (from S-C1 to S-C5).
\item Perceived usefulness. 6 questions in the range 1-7 in which the users must state their agreement level with the proposed questions.
\item Perceived ease of use. Again, 6 questions in the range 1-7 in which the users must state their agreement level with the proposed questions.
\end{enumerate}

Parts (a) and (b) are based on standard Perlman questionnaires~\cite{Perlman1997}, and parts (c) and (d) are based on Davies~\cite{Davis1989} questionnaires. We selected Perlman's because they provide detailed information on the factors that produce the obtained values (3 factors for \textit{Usability} and 5 factors for \textit{Satisfaction}). As these two tests comprise 38 Likert questions and 2 free text questions, and we wanted to avoid a too long questionnaire (up to 50 questions), we selected 2 Davis' brief questionnaires (6 questions each) to check \textit{usefulness} and \textit{easy of use}. In this way we obtained a width scope view of our tool (4 variables) with a reasonable questionnaire size.


The number of participants is a balance between precision and effort. In order to achieve a 90\% confidence level for a given mean with error less than 1\%, it is required to take 15 measurements at least~\cite{Efron1986}.
This assumes intervals based on a normal population distribution for mean, as we consider it is the case.


In the next sections we describe these questionnaires and the results of responses' analysis.

\subsubsection{End-user evaluation of usability and user's satisfaction}
\begin{figure*}
    \centering
    \subfloat[\textit{Usability} part of the questionnaire]{
       \label{fig:usability_results_per_question}
       \includegraphics [width=2.5in]{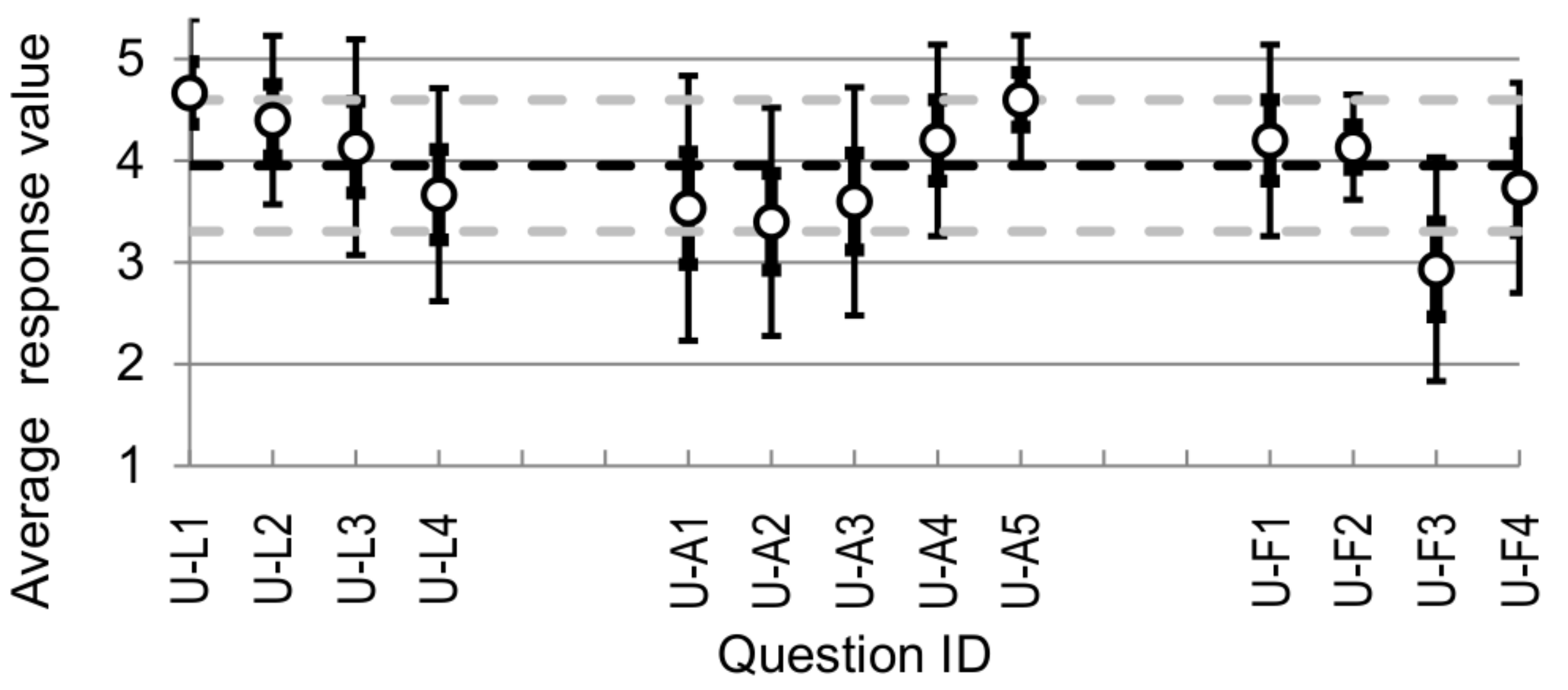}}
    \hspace{0.2cm}
    \subfloat[\textit{Satisfaction} part of the questionnaire]{
       \label{fig:satisfaction_results_per_question}
       \includegraphics [width=3.5in]{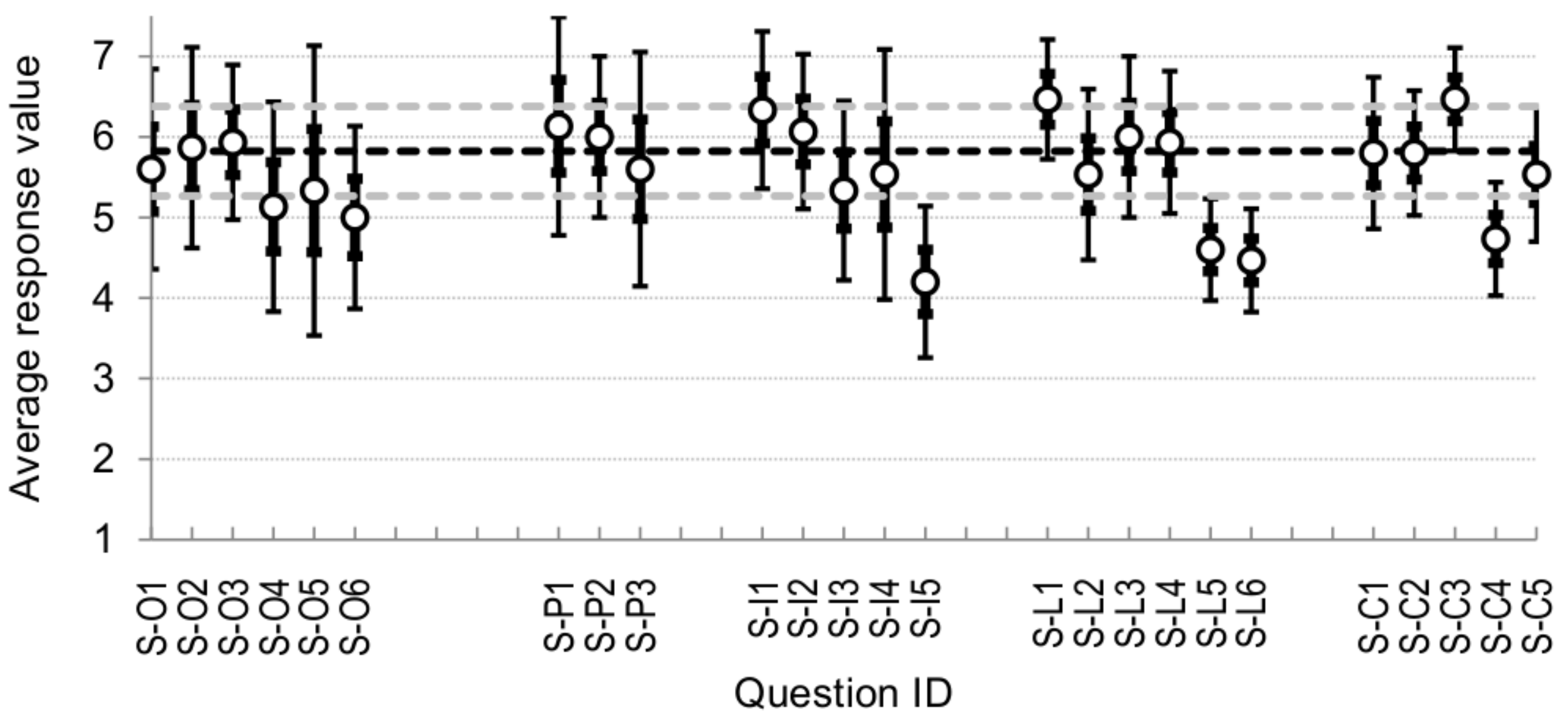}}
    \hspace{0.2cm}
    \subfloat[\textit{Usefulness} part of the questionnaire]{
       \label{fig:perceivedUsability_results_per_question}
       \includegraphics [width=2in]{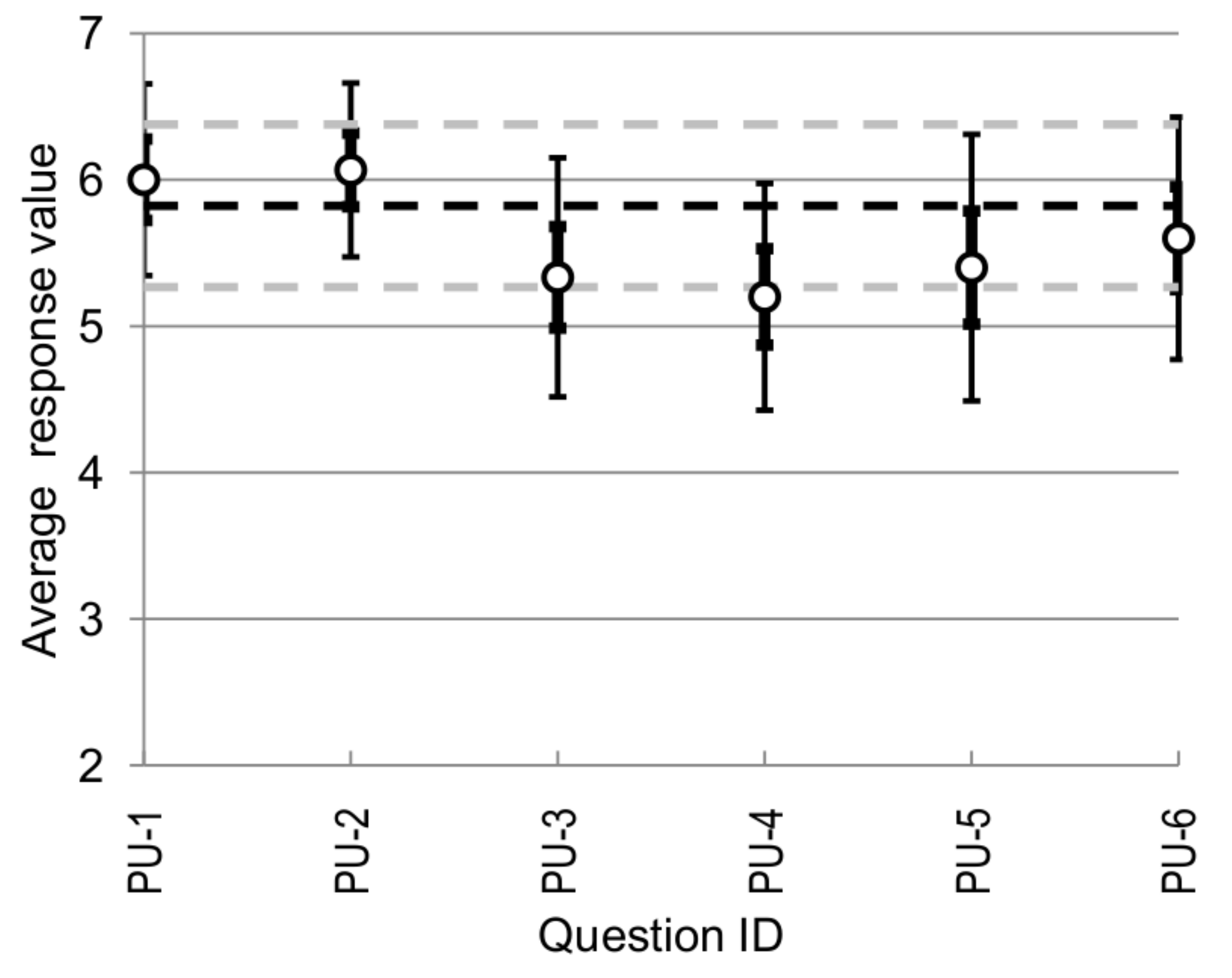}}
    \hspace{0.2cm}
    \subfloat[\textit{Ease of use} part of the questionnaire]{
       \label{fig:perceivedEasy_results_per_question}
       \includegraphics [width=2in]{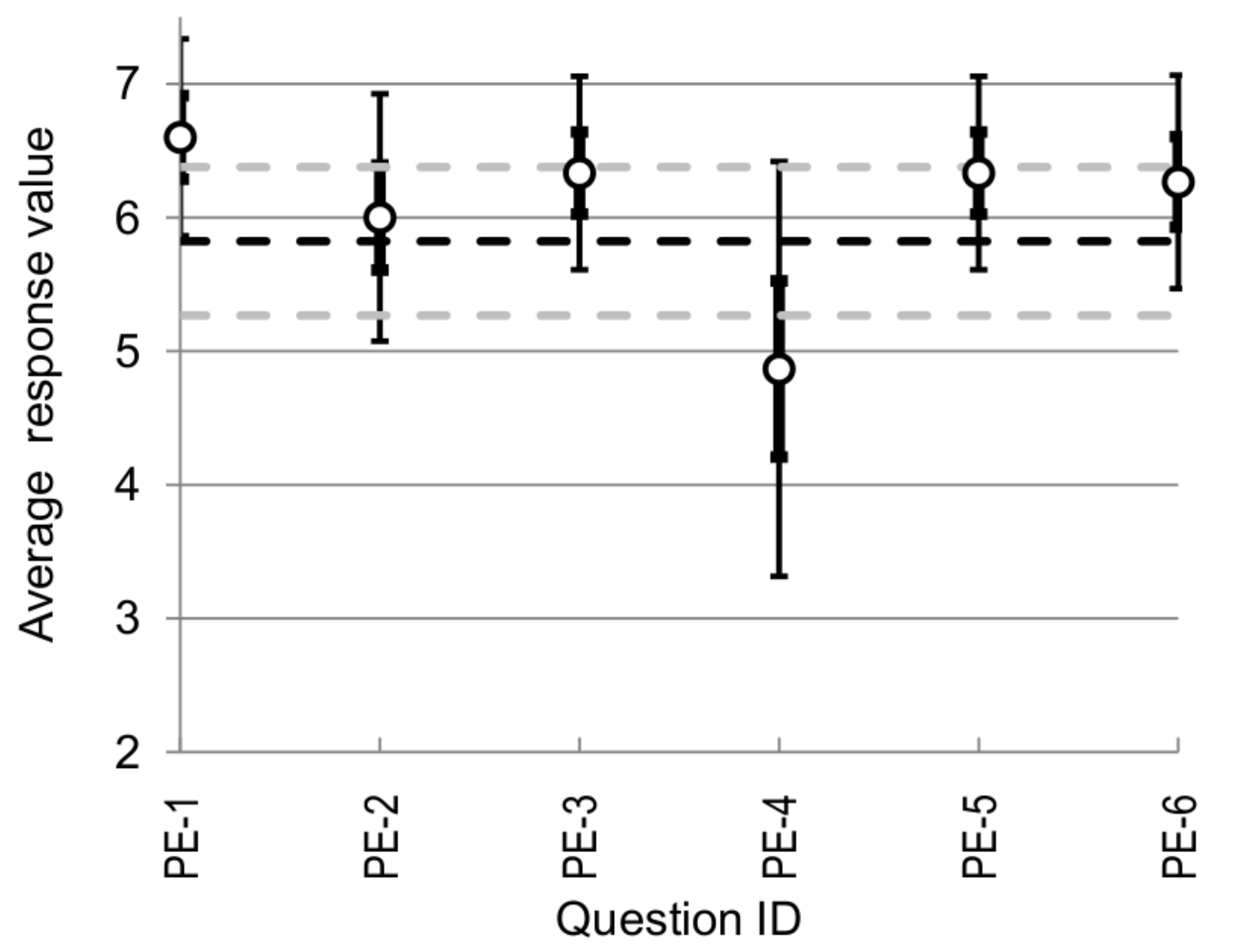}}
    \hspace{0.2cm}
    \caption{Mean value of the user's responses in the questionnaire (circles). Thin bars in each circle show standard deviation. Thick bars show the 90\% confidence interval of the mean. Black dotted lines show the average value of the responses, and grey dotted lines show standard deviation bounds.}
    \label{fig:questionnaire_responses}    
\end{figure*}

The usability of the system was measured by means of the standard test ``Practical Heuristics for Usability Evaluation''~\cite{Perlman1997}. This test includes 13 Likert questions ranging from 1 (bad) to 5 (good), providing a useful measure of the user's perceived usability. Two additional questions are free text questions regarding positive or negatives aspects of the application and its usage. The results of this test are shown in Fig.~\ref{fig:usability_results_per_question}. This figure shows the average value assigned by the participants to each question (thin bars in each circle show standard deviation), the 90\% confidence error bars (thick bars), as well as the global average value (black dotted lines, with value 3.95) and its deviation bounds (grey dotted lines, with standard deviation 0.65). 

The highest values (above 1 standard deviation from the mean) where achieved for questions U-L1 (``Help and documentation'') and U-A5 (``Consistency in the system and to standards''), in both cases with a high consensus (low standard deviation). The highest consensus was for U-F2 (``Prevents errors''). 

Question U-F3 (``Error messages -Diagnose the source and cause of a problem and suggests a solution-'') is below 1 standard deviation, although it has not a big consensus. Between the questions with average value below the mean, we would remark question U-A1 (``Provides a way to preview where to go, what has happened)''). 

In order to evaluate the user's satisfaction, we used a slightly modified version of the standard Perlman test named ``User Interface Satisfaction''~\cite{Chin1988}. The standard version includes 27 questions, but it was reduced to 25 due to overlaps with the usability test described previously. Valid responses to these questions are positive integers ranging from 0 (not satisfied at all) to 7 (completely satisfied). The results are shown in Fig.~\ref{fig:satisfaction_results_per_question}. The average value for user satisfaction was 5.61, with a standard deviation of 0.56.

There are 2 questions with average values above 1 standard deviation of the mean (that is, above the upper dotted grey line), for questions S-L1 (``Learning to operate the system'') and S-C3 (``System tends to be... noisy-quiet''). From here we can claim that this system is easy to learn and very stable.

Concerning questions with lower average values, six questions are below 1 standard deviation of the mean: S-O4 (``application power inadequate-adequate''), SO6 (``application rigid-flexible''), S-I5 (``Error messages unhelpful-helpful''), S-L5 (``Help messages on the screen unhelpful-helpful''), S-L6 (``Supplemental reference materials confusing-clear'') and S-C4 (``Correcting your mistakes never-always''). For the last three there is a very high consensus. However, one question was highly controversial: S-O5 (``overall, application is... dull-stimulating''). 

From these results, and analyzing the free text questions, we identify two kinds of users: users that can understand the recommender explanations, and users that can not (or could not obtain the explanations). The former are more satisfied with the tool, while the later produce lower satisfaction values. In order to enhance the satisfaction, we suggest that the recommender explanation (the recommender analysis) should be available in the main window  instead of the current secondary window. In future versions we also would enhance the explanations to make them more clear. 

%

\subsubsection{End-user evaluation of system perceived usefulness and ease of use}

In order to measure the usefulness and ease of use perceived by end-users, we used the classical test from Davies~\cite{Davis1989}. Each of these two tests comprise 6 Likert questions where participants have to indicate their agreement level concerning each question, from `extremely agree' to `extremely disagree' and 5 levels in between (that is, range 1 to 7). Parts (c) and (d) of figure~\ref{fig:questionnaire_responses} show the average values of the questionnaire concerning perceived usefulness (mean value 5.82, std. dev. 0.42) and ease of use of the system (mean value 5.82, std. dev. 0.58).

The usefulness part shows a uniform agreement on the responses, that is, low standard deviation in question responses and without questions above or below one standard deviation of the mean. The highest value (6.0 over 7.0) was for PU-2 (``Using this tools would improve my job performance'') and the lowest (5.3 over 7.0) was for PU-4 (``Using this tool would enhance my effectiveness on the job''). In both case these high values allow us to conclude that this tool is useful for researchers in tasks such as search of RO's, its reuse, as well as researchers collaboration.

The ``easy of use'' part shows very high values in general, with the exception of question PE-4 (``I find this tool to be flexible to interact with''). This question has a high standard deviation, which is consistent with the existence of two kind of users.

The highest values are for PE-1 (`` Learning to operate this tool is easy for me''), PE-3 (``My interaction with this tool is clear and understandable''), PE-5 (``It is easy for me to become skillful at using this tool'') and PE-6 (``I would find this tool easy to use'').

The free text questions allow getting recommendations from participants. Most evaluators remarked simplicity, intuitiveness, and cleanness as the most relevant perceived benefits. Others pointed that having only one view is good to center the attention focus. Concerning enhancements, some users pointed out that inter-spheres drag\&drop would be a useful feature (as noticed by usability experts).
Most of these recommendations and suggestions have been implemented and are available in the online version of the application. 

\section{Conclusion}
Our approach can help to increase the quality of scientific development in two fundamental ways: i) to facilitate the reuse of previous scientific work related to current research carried out by potentially heterogeneous and delocalized groups of scientists and ii) to reduce the researcher's information overload during their \textit{in silico} experiments by the innovative use of advanced interfaces and recommendation technologies . This is especially relevant given the rise of virtual labs and the need to provide evidences to support the claims made in the literature for validation by the scientific community.

Assisted by usability experts, we have developed CollabSpheres, a web application oriented to assist users in their research. Specifically, this tool allows an exploratory search, assisted by a recommender, aimed at finding research objects relevant to the research activities of the user. This method and tool exploits the semantic descriptions, relations and similarities between research objects and users in order to support advanced search mechanisms.

The experimental evaluation of this tool comprised two parts: the first part was a usability study performed by  5 usability experts of an early prototype. Most experts recommendations were implemented in the final version of the tool. The second part involved 15 participants (researchers) that performed a use case  which covered most of the functionality of the final version tool. These participants filled in a detailed questionnaire (50 Likert questions) which evaluated usability, satisfaction, usefulness and ease of use. The analysis of the responses shows that users perceive high values of simplicity, intuitiveness, and cleanness, as well as this tool increases collaboration and reuse of research objects.

This method and tool also has limitations. The analysis of questionnaire responses shows two kinds of users: users that can understand the explanations made by the recommender system, and users that do not. We will explore enhanced explanation capabilities that convey the rationale of the recommendation to the user in a more effective way.

As follow-up projects derived from this work we could mention 
\begin{enumerate}
\item project EVER-EST~\footnote{See \url{http://ever-est.eu }}, a EU scientific infrastructure effort aimed at creating a Research Object-centric Virtual Research Environment for Earth Science, and their preservation.
\item CollabSpheres has been applied to assist the editorial workflow of publishers, like the American Psychological Association (APA), a publishing body with over 80 journals, 54 interest groups, and 130,000 members, including researchers, educators, clinicians, consultants and students. The CollabSpheres metaphor is enabling reviewers to find capabilities with multiple usability-enhanced and assisted information filtering features. 
\end{enumerate}
These are the first steps in the roadmap for commercialization of CollabSpheres technology in the market of editorial workflow management solutions.

\section*{Acknowledgements} 
This work has been supported by the EU Wf4Ever project (ICT-2009.4.1). Rico, M. wants to thank the support by projects LIDER (EU FP7 project No. 610782) and MINECO's JdC Grant (JCI-2012-12719) and INFRA (UNPM13-4E-1814). G\'{o}mez-P\'{e}rez, JM. wants to thank EU H2020 project EVER-EST (674907, EINFRA-9-2015).


\end{document}